\documentclass[%
 reprint,
 amsmath,amssymb,
 aps, pra,
 longbibliography,
 twocolumn
]{revtex4-1}

\usepackage{graphicx}
\usepackage{dcolumn}
\usepackage{bm}
\usepackage{siunitx}
\usepackage{nicefrac}
\usepackage{mathrsfs}
\usepackage{dsfont}
\usepackage{lipsum}

\usepackage[hidelinks]{hyperref}

\DeclareUnicodeCharacter{2009}{\,}

\begin{document}
	
\renewcommand{\vec}[1]{\mathbf{#1}}
\title{Mitigating higher-band heating in Floquet-Hubbard lattices via two-tone driving}

\author{Yuanning Chen}
\author{Zijie Zhu}
\author{Konrad Viebahn}
\email{viebahnk@phys.ethz.ch}

\affiliation{Institute for Quantum Electronics, ETH Zurich, 8093 Zurich, Switzerland}


\begin{abstract}
Multi-photon resonances to high-lying energy levels represent an unavoidable source of Floquet heating in strongly driven quantum systems.
In this work, we extend the recently developed two-tone approach of `cancelling' multi-photon resonances to shaken lattices in the Hubbard regime. Our experiments show that even for strong lattice shaking the inclusion of a weak second drive leads to cancellation of multi-photon heating resonances. Surprisingly, the optimal cancelling amplitude depends on the Hubbard interaction strength $U$, in qualitative agreement with exact diagonalisation calculations. Our results call for novel analytical approaches to capture the physics of strongly-driven-strongly-interacting many-body systems.
\end{abstract}

\maketitle

\section{Introduction}

Floquet engineering has recently emerged as a powerful tool to realise novel phases of matter that are difficult or impossible to create in their static counterparts~\cite{goldman_periodically_2014,bukov_universal_2015,holthaus_floquet_2016,eckardt_colloquium_2017,basov_towards_2017,oka_floquet_2019,rudner_band_2020,weitenberg_tailoring_2021}.
Examples include the realisation of topological bandstructures~\cite{aidelsburger_realization_2013,miyake_realizing_2013,jotzu_experimental_2014,flaschner_observation_2018,wintersperger_realization_2020,minguzzi_topological_2022,leonard_realization_2023,zhu_reversal_2024}, as well as the dynamic control of charges and correlations in strongly correlated matter~\cite{zenesini_coherent_2009,ma_photon-assisted_2011,meinert_floquet_2016,desbuquois_controlling_2017,gorg_enhancement_2018,messer_floquet_2018,klemmer_floquet-driven_2024}.
In many cases, the most interesting phenomena arise for strong driving: when the rotating-wave approximation breaks down or, equivalently, when higher-order terms in inverse powers of driving frequency become relevant.
However, strong driving can additionally induce unwanted heating to high-lying energy levels~\cite{bakr_orbital_2011,weinberg_multiphoton_2015,strater_interband_2016,reitter_interaction_2017,cabrera-gutierrez_resonant_2019,sun_optimal_2020,song_realizing_2022,zhao_suppression_2022}, such as $p$ bands in optical lattices. 

\begin{figure}[h!]
	\begin{center}
		\includegraphics[width = 0.48\textwidth]{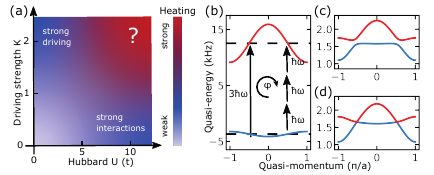}
		\caption{\textbf{The challenge of mitigating higher-band heating in the strongly-driven-strongly-interacting regime (a) and the two-tone cancelling method (b-d).}
            (a), schematic of different Floquet engineering regimes. Strong driving in the non-interacting regime can be solved analytically~\cite{sandholzer_floquet_2022}. Strong interactions in the weak-driving limit have been demonstrated in previous experiments~\cite{viebahn_suppressing_2021}. However, the regime of strong interactions (Hubbard $U \gg$ tunnelling $t$) and strong driving ($K\gtrsim 1$) remains challenging.
            (b-d), Floquet-Bloch spectra of the two-tone cancelling method in a shaken two-band model with an $s$ and a $p$ band.
            The example shows a three-photon resonance as a typical unwanted heating channel, which can be suppressed by a `cancelling' drive that couples the $s$ and $p$ bands directly.
            Single-photon and three-photon couplings can be made to interfere constructively or destructively, depending on the relative phase $\varphi$ and the individual driving strengths $K_1$ and $K_3$.
            (c), a single drive at the fundamental frequency $1\omega$ leads to an $sp$-hybridisation, signaled by the opening of two gaps.
            (d), adding a cancelling drive at $3\omega$ with a specific driving strength and relative phase can cancel out the $1\omega$ coupling and close the gaps.  
		}
		\label{fig:1}
	\end{center}
\end{figure}

In this work, we extend the recently developed `two-tone cancelling' method from refs.~\cite{viebahn_suppressing_2021} and~\cite{sandholzer_floquet_2022} to shaken lattices in the Hubbard regime.
The main challenge is to mitigate higher-band heating when both the drive and the interactions are strong.
In previous observations of two-tone interference, the effects of interactions were studied~\cite{viebahn_suppressing_2021} but the driving strength (lattice amplitude modulation) remained relatively weak.
In the subsequent study with shaken optical lattices~\cite{sandholzer_floquet_2022}, strong driving was reached but the atoms were essentially noninteracting.
Thus, an open question remains, whether strong driving and strong interactions can simultaneously be reached, while maintaining the relevant Floquet physics in the ground band (Fig.~\ref{fig:1}a).

Subjecting a lattice potential to periodic driving gives rise to a Floquet-Bloch band structure~\cite{holthaus_floquet_2016}, in which the energy landscape becomes folded back onto itself in slices of $\hbar \omega$ (the energy of the drive photons).
The folding causes band crossings to emerge between bands which were previously separated by large energy gaps.
In the limit of a vanishing drive, these crossings are exact and no excitations take place, while a finite drive amplitude leads to gap openings (i.e.~resonances) in the Floquet-Bloch spectrum (Fig.~\ref{fig:1}c).
In this context, gap openings correspond to unwanted couplings from the ground band (in which the Floquet physics takes place) to excited bands via absorption of drive photons, which we treat here as heating.
Experimentally, multi-photon resonances are particularly relevant since the fundamental frequency is typically chosen close to the bandwidth of the ground band but smaller than the energy gaps to the excited bands~\cite{messer_floquet_2018}.

The key idea behind two-tone heating cancelling is destructive quantum interference~\cite{viebahn_suppressing_2021} (see also refs.~\cite{schiavoni_phase_2003,zhuang_coherent_2013,niu_excitation_2015,grossert_experimental_2016,gorg_realization_2019,kang_topological_2020,minguzzi_topological_2022,wang_topological_2023,murakami_suppression_2023}).
It can be easiest understood in a two-band model, as illustrated in Fig.~\ref{fig:1}b-c, but it applies to multi-band systems as well~\cite{sandholzer_floquet_2022}.
A multi-photon resonance between $s$ and $p$ bands (due to a strong fundamental drive at $\omega$) is countered by introducing a phase-controlled `cancelling' drive at an integer multiple of $\omega$.
The Rabi-like coupling between $s$ and $p$ resulting from the cancelling drive is chosen to have the same magnitude but opposite sign, compared to the coupling originating from the fundamental drive at $\omega$.
This effectively decouples the two bands, manifested by an exact band crossing (i.e.~gap closing) in case of perfect cancellation, as shown in Fig.~\ref{fig:1}d.
While for amplitude modulation the even harmonics ($2\omega$, $4\omega$) lead to cancelling, lattice shaking usually requires odd harmonics ($3\omega$, $5\omega)$, owing to different symmetries of the spatio-temporal drive pattern~\cite{sandholzer_floquet_2022}.
Practically, one needs to identify the relevant multi-photon resonance which is responsible for heating to a specific higher band;
this can be done by numerically evaluating the Floquet-Bloch spectrum according to ref.~\cite{holthaus_floquet_2016}.
Quantum destructive interference can then be achieved by tuning the driving strength and the relative phase of the `cancelling' drive.
The Floquet-Bloch spectrum only depends on the lattice depth in recoil energies, independently of specific energy scales, such as atomic species and lattice wavelength.
Therfore, the two-tone method is very general and broadly applies to periodically driven optical lattices.

Another advantage of the cancelling method is that the addition of a weak harmonic typically leaves the fundamental-frequency Floquet Hamiltonian unchanged.
For instance, in this work the fundamental drive has a dimensionless strength of $K_1= 1.4$ while the amplitude of the third harmonic is much weaker: $K_3 = 0.07$.
The residual tunnelling renormalisation due to $K_3$ is $\mathcal{J}_0(0.07) = 0.999 \simeq 1$ ($\mathcal{J}_0$ is the zeroth Bessel function of the first kind~\cite{holthaus_floquet_2016}).
Therefore, unless the addition of a harmonic leads to spurious resonances beyond the ones addressed by cancelling, its effect on the ground-band Floquet physics remains negligible.

\section{Experimental setup and measurement of cancelling}

\begin{figure}[htbp]
	\begin{center}
		\includegraphics[width = 0.42\textwidth]{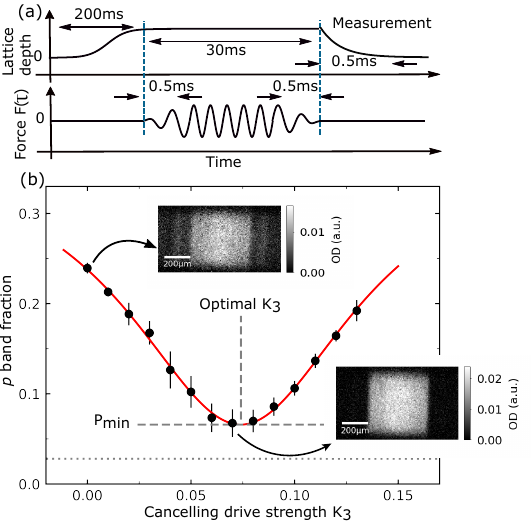}
		\caption{\textbf{Measurement scheme and experimental demonstration of two-tone cancelling.}
        (a) Sketches of lattice depth (top) and forcing amplitude (bottom) as function of time. After the shaking experiment the fraction of atoms excited to the $p$ band is determined via band mapping (see main text).
	(b) In the non-interacting case, a minimum in the measured $p$ band fraction as a function of cancelling drive amplitude $K_3$ demonstrates the ability to effectively cancel a three-photon resonance. The red line denotes the Lorentzian fit of the raw data points (black circles with error bars). The minimum attainable $p$ band fraction ($P_\text{min}$) and the corresponding optimal $K_{3}$ (indicated by the grey dashed lines) are extracted from the fit. The dotted line denotes the minimum detectable $p$ band fraction of 0.028(4).   
		}
		\label{fig:2}
	\end{center}
\end{figure}

Our experiments are performed with ultracold potassium-40 atoms which are loaded into the ground $s$ band of a simple-cubic, three-dimensional optical lattice.
Each lattice direction is formed by retro-reflection of laser light at a wavelength of $\lambda = \SI{1064}{nm}$.
The lattice depths are $[V_\text{X},V_\text{Y},V_\text{Z}] = [5.99(2), 14.95(3), 14.97(5)]\,E_{\text{rec}}$, corresponding to $s$ band tunnelling amplitudes of $[224(1), 29.0(1), 28.9(3)]\, \text{Hz}$ in $x$-, $y$-, and $z$-direction, respectively.
The quantity $E_{\text{rec}}=h^2/(2m \lambda^2)$ denotes one photon recoil of the lattice light, with $h$ being Planck's constant and $m$ being the mass of a potassium-40 atom.
The lattice depths are chosen such that within the duration of Floquet driving, tunnelling to neighouring sites in $y$- and $z$-direction remains small and the system can be described by one-dimensional tubes along $x$.
Prior to lattice loading, which is performed with an $S$-shaped ramp within \SI{200}{ms} (Fig.~\ref{fig:2}a), we set the $s$-wave scattering length between hyperfine states $m_F = -9/2$ and $m_F = -7/2$ ($F = 9/2$) to a value corresponding to a specific Hubbard $U$ in the final lattice.

The position of the retro-reflecting mirror of the $x$-lattice can be modulated in time by a piezo actuator.
The lowest relevant vibrational resonance of the mirror-piezo-mount system is designed to be above \SI{80}{kHz} by choosing a light mirror (quarter-inch diameter), a heavy mount made of tungsten (\SI{213}{g}), and a single-layer piezo (Noliac, NAC2013), which is separated from the tungsten by a \SI{1}{mm}-thick Macor spacer.
The amplitude and phase response of the mechanical system is independently calibrated using a Michelson interferometer with a relative systematic uncertainty of less than $1\%$~\cite{sandholzer_floquet_2022}.
The piezo is driven by a 16bit arbitrary waveform generator (Keysight 33500B) and amplified (PiezoDrive PX200), which ensures the necessary precision in setting the shaking amplitudes $K_1$ and $K_3$.
In the experiment, we `shake' the lattice sinusoidally for \SI{30}{ms} (\mbox{$\simeq$ 42 tunnelling times}), which is equivalent to a periodic forcing of the atoms.
The first and last $\SI{0.5}{ms}$ are used to smoothly ramp the shaking amplitude, as shown in Fig.~\ref{fig:2}a.
Once the lattice shaking is completed, the interactions are switched off within \SI{1}{ms} via a rapid current ramp of the corresponding offset coils to the zero-crossing value of the magnetic Feshbach resonance.
Subsequently the lattice is exponentially ramped to zero (within \SI{0.5}{ms}), followed by absorption imaging, resulting in a measurement of relative band population (`band mapping').
The minimal detectable $p$ band fraction is 0.028(4) due to the smoothness of the visible Brillouin zone boundary resulting from an initial cloud size (before band mapping) of around \SI{25}{\um}.
During the course of the shaking, we do not observe any atom loss.

Lattice shaking results in an inertial force to the atoms, equivalent to an $ac$-electric field for electrons in a solid~\cite{oka_floquet_2019}.
Specifically, the lab-frame Hamiltonian of a single particle in the shaken lattice,
\begin{equation}
\label{eqn:lab_frame}
    \hat{H}_{\text{lab}} = \frac{\hat{P}^2}{2m} + V[\hat{x}-x_{0}(\tau)]~,
\end{equation}
can be transformed into a reference frame that is co-moving with the lattice~\cite{dalibard_reseaux_2013}, leading to
\begin{equation}
\hat{H}_{\text{cm}} = \frac{\hat{P}^2}{2m} + V(\hat{x})-F(\tau)\hat{x},
\label{eqn:co-moving}
\end{equation}
with the time-periodic force $F(\tau)=-m\Ddot{x}_{0}$.

The two-frequency driving waveform studied in this work is 
\begin{equation}
x_{0}(\tau)= A_1 \cos(\omega\tau)+A_{3}\cos(3\omega\tau+\varphi)~,
\end{equation}
which is designed to address a three-photon $p$ band heating channel with its fundamental frequency $1\omega$.
An appropriately optimised $3\omega$ `cancelling' drive can then switch off the heating.
It will be convenient to describe the drive strengths by the dimensionless parameters 
\begin{equation}
K_l = \frac{am\omega_l A_l}{\hbar}~,
\end{equation}
set by the real-space amplitude $A_l$ of the $l$th frequency drive ($a = \lambda/2$).
The parameters $K_l$ describe weak driving for $K\ll 1$ and strong driving for $K \gtrsim 1$, the latter regime being relevant for many Floquet engineering protocols, for example, with $K$ being an argument of Bessel functions~\cite{eckardt_superfluid-insulator_2005,zenesini_coherent_2009,goldman_periodically_2014,bukov_universal_2015}.
In this work, we fix the $1\omega$ driving strength to $K_1 = 1.4$ at an absolute frequency of $\omega_1 = \SI{5300}{Hz}\times 2\pi$, addressing a three-photon 
resonance to the $p$ band (Fig.~\ref{fig:1}b), close to quasimomentum $q = \pm \pi/(2a)$.
The strength of the cancelling drive $K_3$ with $\omega_3 = 3\omega_1 = \SI{15900}{Hz}\times 2\pi$ is varied in the experiment, while the relative phase $\varphi = 0$ is kept fixed.

We first demonstrate cancelling in the non-interacting limit by fixing $U = 0$ and varying the strength of the cancelling drive $K_3$.
The measured $p$ band fraction without cancelling drive ($K_3 = 0$) increases from background level (0.028) to 0.24 within \SI{30}{ms}, as shown in Fig.~\ref{fig:2}b, signalling the presence of significant three-photon coupling in the shaken system.
When the cancelling drive is added, on the contrary, the $p$ band fraction drops down by almost a factor of seven to $P_{\text{min}} = 0.066(5)$ (horizontal dashed line in Fig.~\ref{fig:2}b), close to the background value (dotted line).
This demonstrates the ability to cancel the three-photon resonance by destructive interference.
The minimal attainable $p$ band fraction $P_{\text{min}}$, as well as the optimal cancelling strength $K_{3}$ (0.0741(4) for $U = 0$), is extracted via a lorentzian fit (error bars are obtained via error propagation from the function Nonlinear Curve Fit in Origin).
These measurements are then repeated for different values of Hubbard $U$ (results presented in Section~\ref{sec:results} below).
Heating data without cancelling drive can be found in the Supplemental Material.

\section{Numerical simulations}

In order to capture the relevant physics of the shaking cancelling method, including Hubbard interactions, we set up a minimal model of our experimental system.
Starting from Eq.~\ref{eqn:co-moving}, we add onsite Hubbard interactions and write the whole model in second quantised form: 

\begin{widetext}
\begin{align}
 \label{eqn:hamiltonian}
 \hat{H}(\tau) &= \sum_{j,n,\sigma} \left[E_{n}(\hat{c}_{j,n}^\sigma)^{\dagger}\hat{c}_{j,n}^\sigma+\sum_{k \neq 0}\left(t_{n}^{k}e^{-ik\theta(\tau)}(\hat{c}_{j,n}^\sigma)^{\dagger}\hat{c}_{j+k,n}^\sigma + \text{H.c.}\right)\right] + U \sum_{j,n} \hat{n}^{\uparrow}_{j,n}\hat{n}^{\downarrow}_{j,n} \\
 &- F(\tau)\sum_{j,n'\neq n,\sigma}\left[(\eta_{n,n'}^{0}(\hat{c}_{j,n}^\sigma)^{\dagger}\hat{c}_{j,n'}^\sigma + \text{H.c.})+\sum_{k \neq 0}\left(\eta_{n,n'}^{k}e^{-ik\theta(\tau)}(\hat{c}_{j,n}^\sigma)^{\dagger}\hat{c}_{j+k,n'}^\sigma + \text{H.c.}\right)\right].
 \nonumber
\end{align}
\end{widetext}

This Hamiltonian describes a multi-band Fermi-Hubbard model with periodic forcing, which results in a time-dependent Peierls phase $\theta(\tau)= -\frac{a}{\hbar}\int_{0}^{\tau} F(\tau^{\prime}) d\tau^{\prime}$ ($a = \lambda/2$).
The individual terms are (i) the band on-site energies $E_n = \langle j,n|\hat{H}_{0}|j,n\rangle$, (ii) intra-band tunnelling between $k$th nearest neighbours $t_{n}^{k} =\langle j,n|\hat{H}_{0}|j+k,n\rangle$,
(iii) on-site interactions $U$, (iv) inter-band on-site couplings $\eta_{n,n'}^{0}=\langle j,n|\hat{x}|j,n'\rangle$, and (v) inter-band couplings between $k$th nearest neighbours $\eta_{n,n'}^{k}=\langle j,n|\hat{x}|j+k,n'\rangle$.
The first two terms [(i) and (ii)], are evaluated using the static lattice Hamiltonian $\hat{H}_{0} = \hat{P}^{2}/2m + V_\text{X}\sin^2\left(\frac{2\pi}{\lambda}\hat{x}\right)$.
The latter two terms [(iv) and (v)] and the Peierls phase vanish in the absence of driving.
The usual annihilation and creation operators $\hat{c}_{j,n}^\sigma$ and $(\hat{c}_{j,n}^\sigma)^{\dagger}$, respectively, denote a spin $\sigma$ on lattice site $j$ in band $n$.
In reality, the interaction energy $U$ will also become time-dependent and, in addition, lead to band-dependent variations in $U$ and even couplings between bands~\cite{zhao_suppression_2022}.
However, we neglect these terms for simplicity, as a full evaluation would consist in solving self-consistent Wannier orbitals~\cite{best_interacting_2011}, which goes beyond the scope of this work.

\begin{figure}[htbp]
	\begin{center}
		\includegraphics[width = 0.48\textwidth]{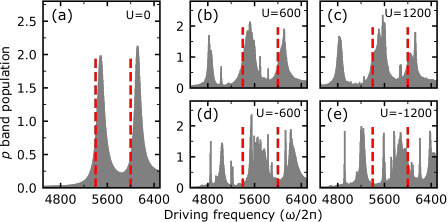}
		\caption{\textbf{Theoretical excitation spectra for a two-band Fermi-Hubbard model under strong driving (single-frequency).}
        The maximum $p$ band population is plotted versus drive frequency for various interaction strengths $U$ (a-e), ranging from \SI{-1200}{Hz} to \SI{+1200}{Hz}.
        The two maxima in the non-interacting case (a, $U = 0$) correspond to two individual coupling points at quasimomenta $q=\pi/2$ (\SI{5490}{Hz}) and $q = 0$ (\SI{6140}{Hz}), due to the small number of fermions in the model ($N_{\uparrow}=2, N_{\downarrow}=2$).
        The dashed red line denotes the frequency window (\SI{5400}{Hz} to \SI{6000}{Hz}) over which the data is averaged for Figs.~\ref{fig:4} and \ref{fig:5}.
        The interacting cases (b-e) exhibit multiple effects, such as the appearance of an additional resonance within the frequency window, narrow resonances throughout the spectrum, as well as a broadening and shifting of the main resonances.
        The frequency window is chosen to contain the main resonance for all values of Hubbard $U$.
		}
  
		\label{fig:3}
	\end{center}
\end{figure}

In the absence of Hubbard interactions, Eq.~\ref{eqn:hamiltonian} can be solved analytically in $q$-space~\cite{sandholzer_floquet_2022}.
As shown in Fig.~\ref{fig:1}b-d, it gives rise to two near-sinusoidal bands, which are coupled via the fundamental drive frequency $1\omega$ (a three-photon resonance close to $q=\pm\pi/(2a)$).
The resonance manifests itself as a gap-opening in the reduced Floquet-Bloch spectrum (Fig.~\ref{fig:1}c).
Adding the cancelling drive at $3\omega$ leads to a gap-closing, which correponds to a de-coupling of $s$ and $p$ bands via destructive interference.

Time-evolving the full Hamiltonian in Eq.~\ref{eqn:hamiltonian} quickly becomes challenging, even for a small number of fermions.
Therefore, we make the following further simplifications:
We work at half filling ($N_{\uparrow}=2, N_{\downarrow}=2$) and restrict the model to two bands, that is, $n \in \{s, p\}$, and four lattice sites $j \in \{1,2,3,4\}$.
Further, we assert periodic boundary conditions and only consider intra-band tunnellings and inter-band couplings up to 2nd order, that is, $k \in \{1,2\}$ in the terms $t_{n}^{k}$ and $\eta_{n,n'}^{k}$.
Including beyond nearest-neighbour terms is particularly important for the $p$ band, which is much more dispersive than the $s$ band~\cite{sandholzer_floquet_2022}.
Exact diagonalisation and time evolution of Eq.~\ref{eqn:hamiltonian} is carried out via the QuSpin python package~\cite{weinberg_quspin_2019} with all system-specific parameters listed in the Supplemental Material.

The numerical simulations are performed by evaluating the many-body ground state of Eq.~\ref{eqn:hamiltonian} and time-evolving this quantum state in the presence of the drive at a fixed frequency (omitting the smooth ramp-up and ramp-down).
The lattice shaking leads to a non-zero density in $p$ band orbitals, which generally grows and/or oscillates in time and also exhibits fast oscillations at the drive frequency (micromotion).
Smoothening the micromotion out by a moving average, the resulting $p$ band population is taken to be the maximum of the time trace during the experimental duration of \SI{30}{ms}.
Repeating the simulation with different drive frequencies gives rise to excitation spectra such as those in Fig.~\ref{fig:3}.
In the non-interacting regime, the presence of single drive at the fundamental frequency $1\omega$ with strength $K_{1}=1.4$ leads to a significant coupling of atoms to the $p$ band, as shown in Fig.~\ref{fig:3}a.
While, compared to the experiment, the appearance of discrete coupling points (maxima in the $p$ band fraction) is an artifact of having only four lattice sites in the system, this small system is still useful, as it allows a straightforward interpretation and sanity-check: In the non-interacting case ($U=0$),
the coupling points correspond to four fermions populating the lowest of the available quasimomenta $q \in \{0, \pm \pi/(2a), \pi/a\}$.
Specifically, the coupling points at $\omega/(2\pi)=\SI{5490}{Hz}$ and $\omega/(2\pi)=\SI{6140}{Hz}$ in Fig.~\ref{fig:3}a correspond to three-photon $s$--$p$ resonances with quasimomenta $q=\pi/(2a)$ and $q=0$, respectively (see also Fig.~\ref{fig:1}b).

\begin{figure}[t]
	\begin{center}
		\includegraphics[width = 0.48\textwidth]{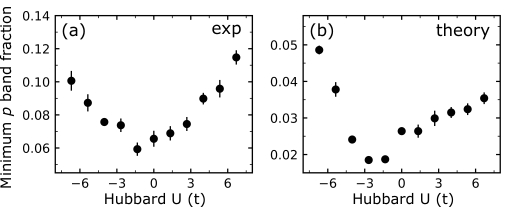}
		\caption{\textbf{Minimum $p$ band fraction ($P_{\text{min}}$) measured as a function of Hubbard $U$.} Panel (a) shows the experimental results, which are determined from Lorentzian fits, such as the one shown in Fig.~\ref{fig:2}b. Panel (b) shows the theoretical results, which are obtained similarly, but using a piecewise linear fit instead of a Lorentzian. Error bars in the experimental data points are propagated from the uncertainties of the `Origin nonlinear fit' function, while error bars in the theory are computed via computing the square root of the diagonal elements of the covariance matrix. The existence of a minimal $p$ band fraction ($P_{\text{min}}$) demonstrates the ability to partially switch off higher-band heating in interacting, strongly-driven Fermi-Hubbard systems. The Hubbard $U$ is given in units of $s$ band tunnelling $t = t^1_s$.
		}
		\label{fig:4}
	\end{center}
\end{figure}

\begin{figure}[t]
	\begin{center}
		\includegraphics[width = 0.48\textwidth]{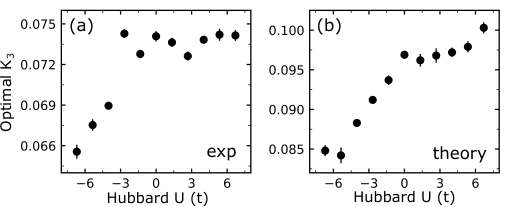}
		\caption{\textbf{Optimal control drive strength ($K_{3}$) measured as function of Hubbard $U$.} The experimental results are plotted in panel (a), showing that for all measured values of Hubbard $U$ an optimal cancelling drive was found. Panel (b) shows the theory results, whose qualitative trend remarkably agrees with the experiment. Quantitative differences between theory and experiment can result from approximations in the theoretical model (see main text), as well as uncontrolled experimental parameters, such as density and temperature.
        Error bars are analogous to Fig.~\ref{fig:4}.
		}
		\label{fig:5}
	\end{center}
\end{figure}

In order to capture the experimental situation in which we expect a continuum of states, rather than discrete coupling points (particularly in the interacting cases), we average the theory simulation over a range of frequencies.
This frequency window is shown in Fig.~\ref{fig:3} as dashed red lines and it is chosen to encompass the relevant resonance at $q = \pi/(2a)$ around $\omega/(2\pi)=\SI{5490}{Hz}$ which can be addressed analytically~\cite{sandholzer_floquet_2022} and `cancelled' in the non-interacting limit (Fig.~\ref{fig:1}d).
The frequency window also serves another purpose:
it allows us to quantify heating and the shaking cancelling efficacy in the interacting regimes where the situation is more intricate than for $U=0$ due to a number of effects.
First, the individual resonances become broader with stronger interactions, as shown in Fig.~\ref{fig:3}b-e.
Second, additional narrow resonances appear which correspond to a combination of interaction-- and shaking--induced processes between the two bands.
Third, the underlying resonances shift to higher frequencies, particularly for strongly attractive interactions (Fig.~\ref{fig:3}e).
In order to account for these effects in the following, all theory simulations shown in Fig.~\ref{fig:4} and Fig.~\ref{fig:5} are averaged over a frequency window ranging from \SI{5400}{Hz} to \SI{6000}{Hz} with a step size of \SI{5}{Hz}.

As in the experiment, we then introduce a cancelling drive at $3\omega$ and repeat the numerical simulations for a range of cancelling drive strengths $K_3$, while keeping all other parameters fixed (relative phase $\varphi = 0$).
For each value of Hubbard $U$, we then obtain a theoretical curve, similar to Fig.~\ref{fig:2}b, from which we extract the optimal $K_{3}$ and the corresponding minimal $p$ band fraction $P_{\text{min}}$.
Contrary to the experimental results, the theoretical curves as a function of $K_{3}$ are fitted with a piecewise linear function (see Supplemental Material), which is likely due to frequency-averaging.
The numerics also exhibits a Lorentzian-like shape in the case of no frequency-averaging (for a single, resonant frequency at $\omega/(2\pi) = \SI{5490}{Hz}$).
We verified that the optimal $K_3$ obtained via frequency averaging agrees with the analytical prediction via solving the Floquet band structure for $U = 0$.

\section{Results: shaking cancelling in the strongly interacting regime}\label{sec:results}

The efficacy of shaking cancelling is evaluated from data such as Fig.~\ref{fig:2}b taken for different values of Hubbard $U$, from which we extract the minimum $p$ band fraction $P_{\text{min}}$ (Fig.~\ref{fig:4}) and optimal cancelling drive strength $K_3$ (Fig.~\ref{fig:5}).
The existence of a value of $P_{\text{min}}$, which signals a minimum of $p$ band atoms as a function of $K_3$, demonstrates the ability to reduce higher-band heating in all measured interaction regimes up to $\vert U \vert < 7t$, including strongly attractive ($U < 0$) and strongly repulsive ($U>0$) interactions.
Taking the value of $P_{\text{min}}$ as a proxy for the quality of this effect, the cancelling becomes less efficient for stronger interactions, independently of the sign of $U$.
This qualitative effect is visible in theory (Fig.~\ref{fig:4}b) and experiment (Fig.~\ref{fig:4}a), although it is more pronounced in the experimental data.
We also extract the width of the cancelling minima such as Fig.~\ref{fig:2} and we find that it does not depend on $U$.
The theoretical model fails to capture the absolute value of $P_{\text{min}}$, which suggests that additional effects could play a role here, such as harmonic trapping, temperature, varying density, and higher bands.
These effects may lead to a broad background heating, which is independent of two-tone cancelling, causing the theory to consistently underestimate $P_{\text{min}}$ compared to the experimental value.
We leave refinements of the theoretical description for future work, such as going beyond the two-band model, including harmonic trapping, and increasing the system size.

Notably, the best cancelling effect occurs for small negative $U$, in both theory and experiment (although it is within error bars in the latter case).
The fact the heating-mitigation method works best in the non- or weakly-interacting regime agrees with the intuition that cancelling relies on distinct excitation channels which can be addressed individually.
Strong interactions lead to a broadening of the excitation spectra (Fig.~\ref{fig:3}), which means that more and more excitation pathways appear.
Therefore, heating suppression becomes less straightforward in strongly interacting systems.
Nevertheless, we observe the cancelling effect for all interaction strengths tested here, which is an important result of our work.

Another key observation is the clear dependence of the optimal cancelling strength $K_3$ as a function of Hubbard $U$ in Fig.~\ref{fig:5}.
While attractive interactions lead to a smaller value of optimal $K_3$, repulsive interactions favour a slightly larger $K_3$.
This result is unexpected and nontrivial, as the cancelling effect itself has so far only been analytically predicted for non-interacting atoms~\cite{sandholzer_floquet_2022}.
Our observations show that strong interparticle interactions can lead to significant changes in the cancelling mechanism, suggesting that cancelling can should optimised for every specific drive strength individually.

The theory simulation is able to reproduce the trend of optimal $K_3$ as a function of Hubbard $U$, supporting the experimental observation.
Quantitative differences in the predicted value of optimal $K_3$ between theory and experiment can result from approximations made in the model (neglecting higher-order terms, restricting to two bands, finite-size effects), as well as parameters in the experiment which are not completely controlled, such as density and temperature.
Nevertheless, the theory-experiment comparison provides an important cross-validation, which is not trivial for strongly interacting, strongly driven many-body systems~\cite{sandholzer_quantum_2019}

\section{Conclusion}

In conclusion, we have experimentally and theoretically investigated the performance of two-tone cancelling in strongly-driven-strongly-interacting Fermi-Hubbard systems.
While interactions lead to a broadening of the excitation spectrum, the two-tone method nevertheless improves the ground-state coherence for all interaction strengths tested in this work.
Interestingly, we observed a clear dependence of optimal cancelling parameters on the interaction strength.
Overall, the experimental results are supported by exact diagonalisation calculations in small systems.

We emphasise that the experimental realisation of two-frequency driving is not limited to cold-atom or photonic platforms, but applies in general to Floquet-driven matter.
For instance, frequency-doubled laser light constitutes a natural implementation of $1\omega$--$2\omega$ driving schemes in the optical domain~\cite{heide_optical_2021}.
The underlying single-particle Hamiltonian (Eq.~\ref{eqn:co-moving}) is equivalent to an oscillating electric field $e E(\tau) = F(\tau)$ coupled to free electrons in a conductor ($e$ denotes an electron charge).
Thus, our results could also be applied to condensed matter and quantum optics.
Moreover, the observation of an unexpected shift in the optimal cancelling parameters as a function of interaction strength calls for novel analytical approaches in driven Hubbard models.
Possible strategies towards making ab-initio predictions for higher-band heating mitigation include the Floquet-Fermi-Golden rule~\cite{bilitewski_scattering_2015,ikeda_fermis_2021}, Floquet-Boltzmann equations~\cite{genske_floquet-boltzmann_2015}, semiclassical treatments~\cite{li_floquet_2019}, (nonequilibrium) dynamical mean-field theory~\cite{aoki_nonequilibrium_2014,qin_nonequilibrium_2018,sandholzer_quantum_2019,murakami_suppression_2023}, and others~\cite{bukov_heating_2016,mallayya_heating_2019,mori_heating_2022,queisser_higher-harmonic_2024}.
Another avenue for improving ground state coherence in the presence of interactions consists in increasing the number of cancelling control parameters, for instance via adding many harmonics to the drive waveform, and optimising their amplitudes and phases algorithmically~\cite{castro_floquet_2022}.

\section*{Acknowledgments}
We would like to thank Tilman Esslinger for supporting this project.
We acknowledge funding by the Swiss National Science Foundation (Grants No.~182650, No.~212168, and NCCR-QSIT), European Research Council advanced grant TransQ (Grant No.~742579), as well as Quantera dynamite PCI2022 132919. K.V.~acknowledges support by the ETH fellowship programme.


%

\clearpage
\newpage

\setcounter{figure}{0} 
\renewcommand\thefigure{A\arabic{figure}}

\section*{Supplemental material}

The Supplement contains additional theoretical and experimental data, as well as the numerical values of system-specific parameters.

{\renewcommand{\arraystretch}{1.2}
\begin{table}[ht]
    \centering
    \begin{tabular}{ccc}
    \hline\hline
    Parameter & Value & Unit\\
    \hline
    $E_s/h$ & $-3697.0$ & Hz\\
    $E_p/h$ & $12513.4$ & Hz\\
    $t_s^{1}/h$ & $-224.0$ & Hz\\
    $t_s^{2}/h$ & $8.7$ & Hz\\
    $t_p^{1}/h$ & $1694.1$ & Hz\\
    $t_p^{2}/h$ & $278.4$ & Hz\\
    $\eta_{s,p}^{0}$ & $0.1641$ & N/A\\
    $\eta_{s,p}^{1}$ & $-0.0314$ & N/A\\
    $\eta_{s,p}^{2}$ & $0.0033$ & N/A\\
    \hline
    \end{tabular}
    \caption{\textbf{Numerical values of system-specific parameters in the Hamiltonian (Eq.~\ref{eqn:hamiltonian}).}}
    \label{tab:parameters}
\end{table}
}

\begin{figure}[htbp]
	\begin{center}
		\includegraphics[width = 0.30\textwidth]{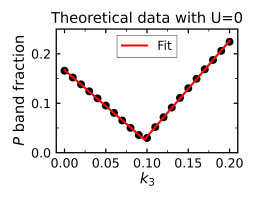}
		\caption{Determining the optimal $K_3$ in the numerical simulation in analogy to Fig.~\ref{fig:2}b.
		}
		\label{fig:a1}
	\end{center}
\end{figure}

\begin{figure}[htbp]
	\begin{center}
		\includegraphics[width = 0.48\textwidth]{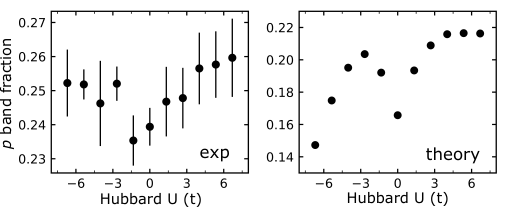}
		\caption{\textbf{Single-frequency heating in the presence of Hubbard interactions.} The $p$ band fraction is measured after \SI{30}{ms} of single-frequency lattice shaking (no cancelling) as function of Hubbard $U$ in the experiment (left) and in numerics (right). The reduction of $p$ band fraction on the attractive side in the numerics is likely an artifact of the frequency window (see main text).
		}
		\label{fig:a2}
	\end{center}
\end{figure}

\end{document}